# Accelerated Lanthanide Intercalation into Graphite Catalyzed by Na

*Akira Iyo,[1*] Hiroshi Fujihisa,[1] Yoshito Gotoh,[1] Shigeyuki Ishida,[1] Hiroshi Eisaki[1], Hiraku Ogino[1], Kenji Kawashima[1,2]*

[1]National Institute of Advanced Industrial Science and Technology (AIST), Tsukuba, Ibaraki 305-8568, Japan

[2] IMRA JAPAN Co., Ltd., Kariya, Aichi 448-8650, Japan

**ABSTRACT:** Lanthanides (*Ln*) are notoriously difficult to intercalate into graphite. We investigated the possibility of using Na to catalyze the formation of *Ln*-intercalated graphite and successfully synthesized $LnC_6$ (*Ln* = Sm, Eu, and Yb) significantly rapidly in high yields. The synthesis process involves the formation of the reaction intermediate $NaC_x$, through the mixing of Na and C, which subsequently reacts with *Ln* upon heating to form $LnC_6$. Well-sintered $LnC_6$ pellets with low residual Na concentrations (*Ln*:Na ≈ 98:2) were fabricated by the two-step method. The pellets enabled the evaluation of $LnC_6$ by powder X-ray diffraction and electrical resistivity measurements. This study highlights the versatility of the Na-catalyzed method and lays the foundation for the rapid mass production of $LnC_6$, with potential applications in superconducting and rechargeable battery materials.

1. INTRODUCTION

Graphite intercalation compounds (GICs) are formed by the insertion of atoms or molecules (intercalates) into graphite, consisting of stacks of graphene sheets. Alkali metals ($A_M$), alkaline earth metals ($A_E$), and lanthanoids (*Ln*) are known as typical donor-type intercalates. Among these, $A_M$ can easily intercalate into the graphite, allowing for the synthesis of $A_M$-GIC bulk samples by the vapor-phase method.[1] In contrast, the intercalation of $A_E$ or $Ln$ is not as facile. Bulk samples of $A_E$-GICs ($A_E C_6$) have been synthesized by the molten alloy[2,3] and molten salt[4,5] methods. However, these methods commonly require time-consuming heat treatment at a temperature ($T$) of 350 °C for approximately a week.

Recently, the use of Na as a catalyst was reported to drastically accelerate the formation of an $A_E$-GIC ($A_E C_6$).[6,7] Bulk samples of $A_E C_6$ can be produced by simply mixing $A_E$ and C powders with Na at room temperature, followed by heat treatment at a relatively low $T$ (250 °C) for just a few hours. This Na-catalyzed method is characterized by the formation of the reaction intermediate $NaC_x$ (stage 6–8 Na-GIC with $x$ of 48–64)[8,9] via the mixing of Na with C. Owing to its instability,[10,11] $NaC_x$ transforms to the more stable $A_E C_6$ by reacting with $A_E$ upon heating.

The intercalation of lanthanides ($Ln$) into graphite is even more challenging. In 1980, Makrini et al. attempted the synthesis of $Ln$-GICs ($LnC_6$) for $Ln$ = Sm, Eu, Tm, and Yb.[12] However, even with an extended period of heat treatment (≈ 20 d), the yield reached a maximum of only 25% for $EuC_6$. Hagiwara et al. reported the formation of $LnC_6$ ($Ln$ = Nd, Sm, Dy, Er, and Yb) by immersing highly oriented pyrolytic graphite (HOPG) in a eutectic molten solution of LiCl-KCl containing $Ln$ and $LnCl_3$, followed by 4–5 days of heat treatment.[13] However, their paper only presented the lattice constants, and lacked descriptions of the X-ray diffraction (XRD) patterns and yields of $LnC_6$. Despite subsequent attempts to synthesize $YbC_6$ using either the vapor-phase[14,15] or the molten salt[16] method, $YbC_6$ was limited to forming on the surface of HOPG. Only $EuC_6$ was synthesized in the bulk form via heat treatment at 350 °C for 10 days using either the molten salt[17] or the molten alloy[18] method.

The above-mentioned unsatisfactory yields and time-consuming processes have limited the research and development of $LnC_6$ for potential applications such as superconducting materials,[14] hydrogen storage materials,[19] and anode materials for secondary batteries.[20,21] This motivated us to devise a more efficient method for the synthesis of bulk quantities of $LnC_6$ by applying the Na-catalyzed method. Initially, the potential for Na-catalyzed $Ln$-GIC formation was comprehensively investigated for $Ln$ = La, Ce, Pr, Nd, Sm, Eu, Gd, Tb, Dy, Ho, Er, Tm Yb, and Lu. As a result, the three $LnC_6$ ($Ln$ = Sm, Eu, and Yb) were found to form significantly rapidly in high yields although the catalytic effect of Na did not lead to the formation of $Ln$-GICs that had not been synthesized previously. In this paper, we report the details of the Na-catalyzed synthesis of $LnC_6$ ($Ln$ = Sm,

Eu, and Yb) and an experimental study on their formation process. Furthermore, we fabricated $Ln$C$_6$ pellets with low residual Na concentrations and evaluated their physical properties via powder XRD and electrical resistivity measurements. This paper provides a foundation for the efficient production of $Ln$C$_6$ and encourages the research and development of $Ln$C$_6$ for various applications.

## 2. EXPERIMENTAL SECTION

### 2.1 Sample Preparation

Graphite powder (Furuuchi Chemical Corp., 99.99%, -200 mesh) was employed as the host for intercalation. $Ln$ lumps (Furuuchi Chemical, 99.5%) were filed down to powder form prior to use. The soft metal Na was used by directly cutting off a lump (Furuuchi Chemical, 99.9%). In a previous study,[6,7] the materials of Na, $A_E$, and C were mixed simultaneously. In this study, $Ln$ powder and Na were first mixed (kneaded) for ~15 min using a mortar and pestle to finely disperse the $Ln$ particles in Na. C powder was then added to this mixture of $Ln$ and Na (the materials weighed ~ 0.2 g in total), whereupon mixing was continued for an additional 15 min. We used a mixing ratio of $Ln$:C:Na = 1:6:2, where the mixing ratio of Na was determined in another study that synthesized CaC$_6$ under conditions similar to those of this study.[7] The resulting sample was pressed into a pellet. The pellet was sealed in either stainless steel (SST) or quartz tubes. Owing to the reactivity of Na with quartz, SST tubes were preferentially used for heating above 275 °C for more than 2 h. The detailed heat-treatment conditions of the samples are described in the following sections. As the $Ln$, Na and $Ln$C$_6$ samples are unstable in air, they were handled in an Ar-filled glove box.

### 2.2 Measurements

XRD experiments were conducted using Cu$K\alpha$ radiation on a Rigaku Ultima IV instrument with an airtight attachment to prevent the sample from being exposed to air. Samples without Na reduction could not be powdered owing to their high viscosity, which inevitably led to the $c$-axis orientation of $Ln$C$_6$ crystals at the sample surface when they were placed on the sample plate for XRD measurements. In contrast, samples from which most of the Na had been removed could be powdered, enabling the collection of XRD patterns with a suppressed orientation.

The compositions of the samples were analyzed using an energy-dispersive X-ray spectrometer (Oxford, SwiftED3000) equipped with an electron microscope (Hitachi High Technologies, TM-3000). The dependence on $T$ of the resistivity ($\rho$) was measured utilizing the four-terminal method on a physical property measuring system (Quantum Design, PPMS). The electrodes were affixed to the sample in a glove box using Ag paste (DuPont 4922N). Once the Ag paste had dried, the sample was covered with Apiezon N grease to prevent exposure to air. The dependence of the magnetization ($M$) on $T$ was measured using a magnetic property measurement system (Quantum Design, PPMS).

## 3. RESULTS AND DISCUSSION

### 3.1 Control experiments to Verify Catalytic Effect of Na

Prior to the Na-catalyzed synthesis of $Ln$C$_6$, control experiments were conducted to determine the extent to which $Ln$C$_6$ ($Ln$ = Sm, Eu, and Yb) forms without the addition of Na. $Ln$ and C powders were mixed in a molar ratio $Ln$:C of 1:6, pelletized, and heat-treated at 275 °C for 6 h in SST tubes. The powder XRD patterns of the resulting samples are shown in Figure 1. For $Ln$ = Eu, although clear diffraction peaks from EuC$_6$ were observed, the majority of the sample still consisted of unreacted Eu and C. For $Ln$ = Yb, only tiny diffraction peaks from YbC$_6$ were detected, and SmC$_6$, known for its challenging synthesis as mentioned in the literature,[12] did not form at all.

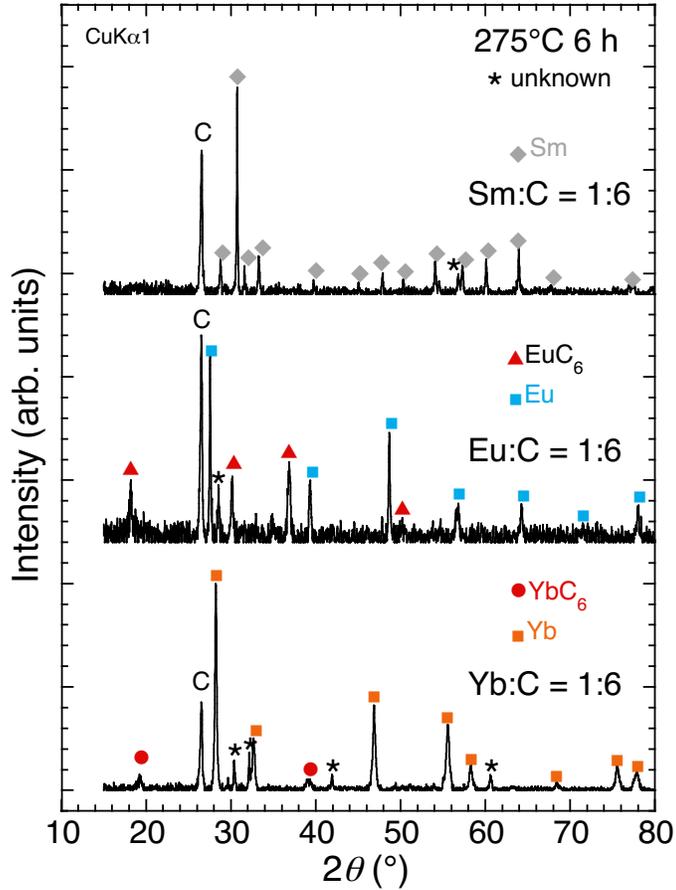

**Figure 1.** Powder XRD patterns of samples prepared without the addition of Na. The experiments were performed as a control to examine the catalytic effect of Na on the formation of $Ln$C$_6$ ($Ln$ = Sm, Eu, and Yb).

Next, we investigated the catalytic effect of Na on the formation of SmC$_6$ as a function of the heat-treatment temperature ($T_h$). Samples with a Sm:C:Na mixing ratio of 1:6:2 were pelletized and sealed in quartz tubes, followed by heat treatment at $T_h$ of 150, 225, and 275 °C for 2 h. The XRD patterns of the samples are presented in Figure 2 as a function of $T_h$. In the as-mixed sample (i.e., without heat treatment), only NaC$_x$ was formed by the reaction of C and Na, whereas Sm remained unreacted. The formation of NaC$_x$ by mixing alone has also been observed in the synthesis of Li-GIC and CaC$_6$.[6,7] At $T_h$ = 150 °C, SmC$_6$ did not form, and instead, an unknown phase (presumed to be a Na-Sm binary compound) was generated as denoted by asterisks. At $T_h$ = 225 °C, the emergence of the diffraction peaks confirmed the formation of SmC$_6$. However, the sample still contained a significant fraction of the unreacted Sm and the unknown phase. At $T_h$ = 275 °C, the yield of SmC$_6$ increased rapidly, accompanied by a marked decrease in the peak intensities of the unknown phase and $NaC_x$. These results, compared to the results of the control experiment, clearly demonstrate that Na has a strong catalytic effect on the formation of $SmC_6$, with $NaC_x$ acting as a reaction intermediate of the catalyst.

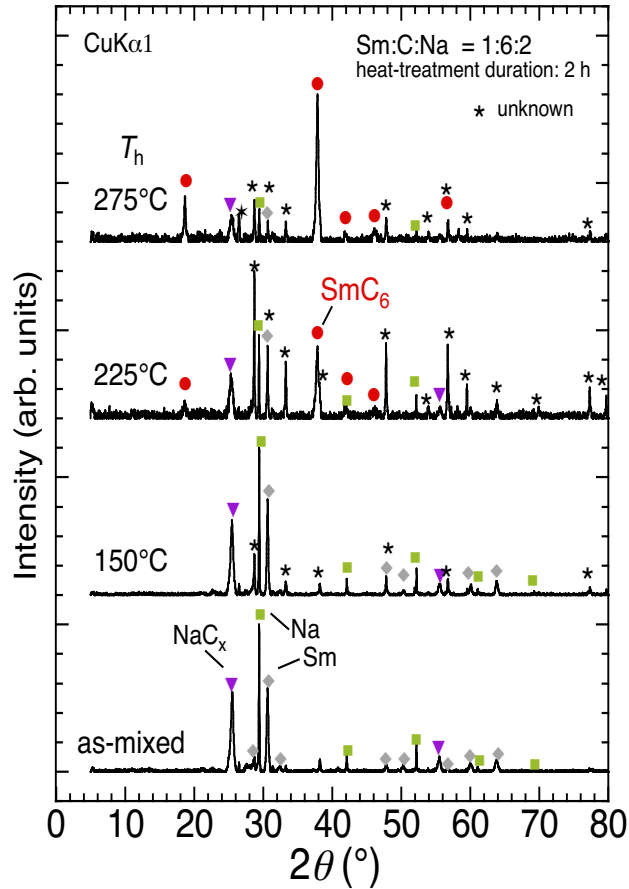

**Figure 2.** XRD patterns of samples (Sm:C:Na = 1:6:2) as a function of the heat-treatment temperature ($T_h$).

Next, we investigated the catalytic effect of Na as a function of the heat-treatment duration ($D_h$) for $YbC_6$. Samples with a Yb:C:Na mixing ratio of 1:6:2 were pelletized and placed in SST tubes, followed by heat treatment at 275 °C for $D_h$ of 0.25, 1, and 6 h. The acquired XRD patterns are shown in Figure 3. The XRD pattern of the as-mixed sample, shown by the dashed curve, primarily consisted of the diffraction peaks of $NaC_x$, Yb, and Na. With a heat treatment of only $D_h = 0.25$ h, a high-intensity diffraction peak of $YbC_6$ appeared at $2\theta \approx 39°$, accompanied by a significant decrease in the intensity of the $NaC_x$ and Yb peaks. This is clearly indicative of the formation of YbC$_6$ via the reaction between NaC$_x$ and Yb. By increasing $D_h$ to 1 h and then 6 h, the formation of YbC$_6$ progressed steadily. Thus, the catalytic effect of Na drastically accelerated the formation of $Ln$C$_6$.

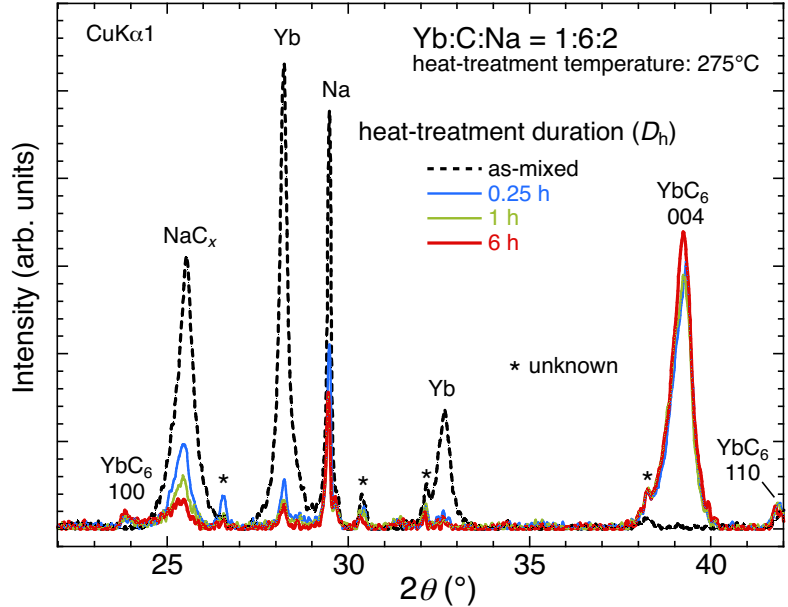

**Figure 3.** XRD patterns of the samples (Yb:C:Na = 1:6:2) as a function of the heat-treatment duration ($D_h$).

### 3.2 Fabrication of Pellets with Reduced Residual Na

Samples prepared by the Na- catalyzed method inevitably result in the mixture of $Ln$C$_6$ and Na. For the potential application of $Ln$C$_6$, the residual Na in the sample needs to be minimized. Samples with a $Ln$:C:Na mixing ratio of 1:6:2 ($Ln$ = Sm, Eu, and Yb) were pressed into pellets (the weight of each pellet was approximately 200 mg). The samples were heated at 275 °C in SST tubes for a total duration of 12 h for $Ln$ = Sm and Yb, and 4 h for $Ln$ = Eu. The heat treatment was interrupted once for an intermediate mixing of the sample to improve the yield and homogeneity of $Ln$C$_6$.

Subsequently, using a two-step Na reduction process,[7] polycrystalline pellets with low residual Na concentrations were fabricated. The $Ln$:Na composition ratio of the pellets was measured to be ≈ 98:2. Because the intercalated $Ln$ is divalent, as is the case for $A_E$,[22] the $Ln$C$_6$ pellets have a light golden color similar to that of $A_E$C$_6$,[6] as shown in Figure 4. The densities of the $Ln$C$_6$ pellets ($Ln$ = Sm, Eu, and Yb) were 3.66, 3.69, and 4.29 g/cm$^3$, respectively, approximately 77% of the theoretical density. Note that this is the first time that bulk samples of $SmC_6$ and $YbC_6$ were successfully fabricated.

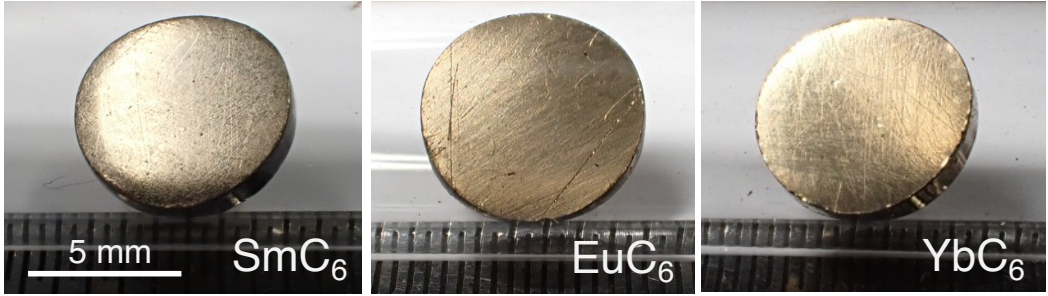

**Figure 4**. Polycrystalline $LnC_6$ pellets ($Ln$ = Sm, Eu, and Yb) with low residual Na concentrations ($Ln$:Na ≈ 98:2).

The XRD patterns of the $LnC_6$ powders ($Ln$ = Sm, Eu, and Yb) acquired by grinding the pellets are depicted in Figure 5a. By carefully placing the powders on the sample plates to minimize the preferred orientation of $LnC_6$ crystals, XRD patterns containing peaks with diffraction indices other than $00l$ were obtained.

The diffraction peaks of $LnC_6$ ($Ln$ = Sm and Eu) could be indexed well by assuming the same hexagonal system ($P6_3/mmc$) as $BaC_6$ and $SrC_6$. The appearance of $h00$ diffraction peaks is a hallmark of the hexagonal crystal system. The diffraction peaks of $YbC_6$ could also be indexed basically by assuming the same hexagonal system. However, the $00l$ diffraction peaks were broad, and weak diffraction components that could be explained by assuming a rhombohedral crystal system ($R$-$3m$) appeared (denoted by arrows). This suggests that the rhombohedral stacking sequence of Yb and graphene layers ($A\alpha A\beta A\gamma A\alpha$) coexisted with the hexagonal stacking sequence ($A\alpha A\beta A\alpha A\beta$).[12] Note that $CaC_6$, synthesized via the Na-catalyzed method, is suggested to have an inverse stacking sequence disorder ($A\alpha A\beta A\gamma A\alpha$ stacking sequence combined with $A\alpha A\beta A\alpha A\beta$).[7]

Table 1 presents the $a$- and $c$-axis lengths of unit cells, interlayer distances ($d_s$), and unit cell volumes ($V$) of $LnC_6$ derived from the analysis of the powder XRD patterns. The interlayer distance refers to the spacing between adjacent graphene layers, corresponding to half the $c$-axis length for $LnC_6$. Our values are in close agreement with those in the literature.[12] The $d_s$ values of $LnC_6$ and $A_EC_6$ obtained by the Na-catalyzed method are plotted against the ionic radii of divalent lanthanides ($Ln^{2+}$) and alkaline earth metals ($A_E^{2+}$) in Figure 5b.[23] The relationship between the ionic radii and $d_s$ is approximately linear. Despite the close proximity of $YbC_6$ and $CaC_6$, the structural boundary between rhombohedral and hexagonal systems exists between them. Therefore, it seems natural for two different stacking sequences to intermingle in YbC$_6$ and CaC$_6$ as a result of the Na-catalyzed rapid synthesis.

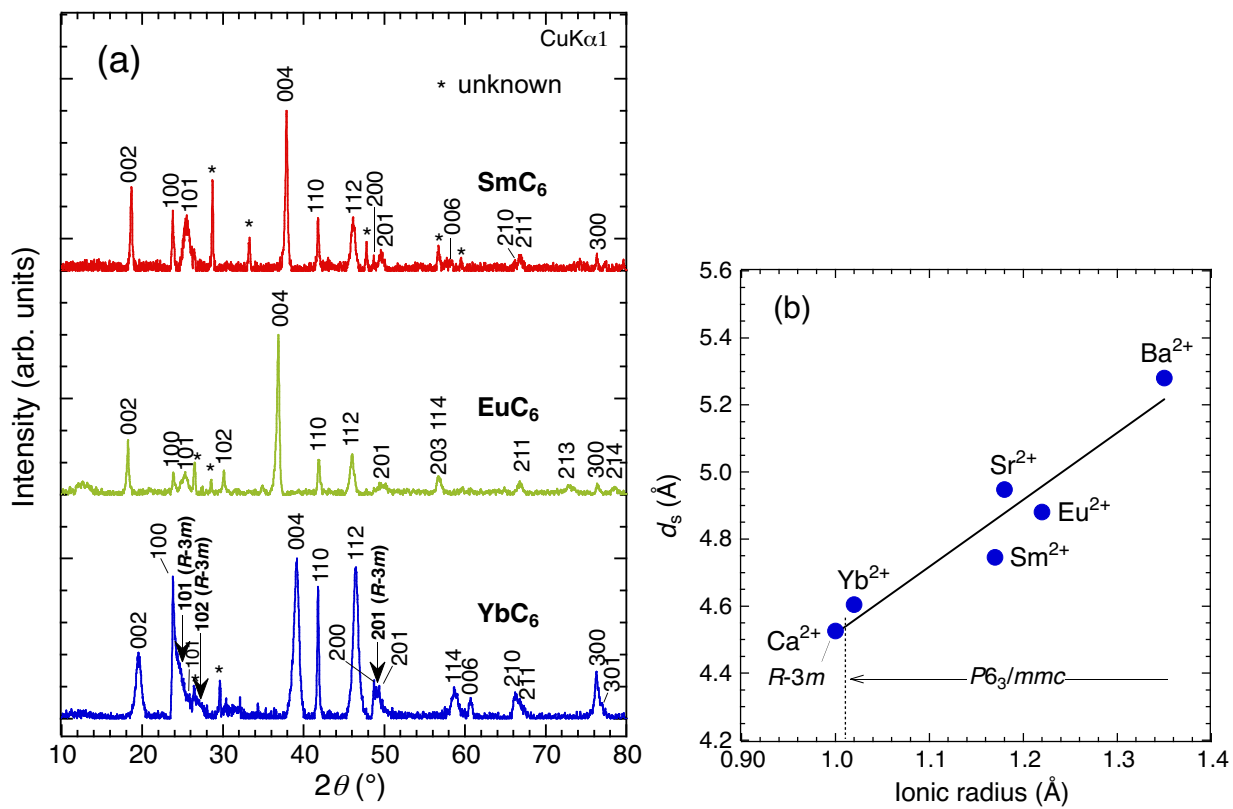

**Figure 5.** (a) Powder XRD patterns of $Ln$C$_6$ ($Ln$ = Sm, Eu, and Yb). The diffraction peaks are indexed assuming a hexagonal crystal system ($P6_3/mmc$). For YbC$_6$, the diffraction angles and indices (hexagonal setting) that appear when the rhombohedral crystal system ($R$-$3m$) is assumed are also indicated. (b) Correlation between the interlayer distance ($d_s$) and the ionic radius of divalent $Ln$ and $A_E$ for $Ln$C$_6$ and $A_E$C$_6$.

**Table 1.** $a$- and $c$-axis lengths of unit cells, interlayer distances ($d_s$), and unit cell volumes ($V$) of $Ln$C$_6$ synthesized in this study, along with values reported in the literature.[12]

| GICs | $a$ (Å) | $c$ (Å) | $d_s$ (Å) | $V$ (Å$^3$) | Refs. |
|---|---|---|---|---|---|
| SmC$_6$ | 4.319(1) | 9.490(2) | 4.745(1) | 153.3(1) | this study |
|  | – | ~9.40 | ~4.70 | – | 12 |
| EuC$_6$ | 4.319(1) | 9.761(1) | 4.881(1) | 157.7(1) | this study |
|  | 4.314(3) | 9.745(8) | 4.873(4) | 157.2 | 12 |
| YbC$_6$ | 4.320(1) | 9.207(1) | 4.604(1) | 148.8(1) | this study |
|  | 4.320(4) | 9.147(4) | 4.574(2) | 147.8 | 12 |

### 3.3 Physical Properties of $Ln$C$_6$

The low residual Na concentration in the $Ln$C$_6$ pellets minimizes the effect of Na on transport properties. The dependence of the resistivity $\rho$ on $T$ for $Ln$C$_6$ ($Ln$ = Sm, Eu, and Yb) is depicted in Figure 6a. All samples exhibit metallic behavior, with $\rho$ in the range of 0.3–0.4 mΩ cm at 300 K; however, their behaviors diverge at lower $T$. While SmC$_6$ did not exhibit anomalous behavior above 2 K, a bump corresponding to a magnetic phase transition was detected at 39.0 K in EuC$_6$.[22,24]

For YbC$_6$, an abrupt resistive drop due to the occurrence of superconductivity was observed at a midpoint (onset) transition temperature of $T_c^{mid}$ = 5.1 K ($T_c^{onset}$ = 6.0 K), as indicated in the inset. The superconductivity was also confirmed by magnetization measurements (Figure 6b). The nearly 100% shielding volume fraction ($4\pi M/H \approx -1$ at 2 K in the zero-field-cooling (ZFC) curve) indicates that the sample is well-sintered. The YbC$_6$ sample prepared in this study exhibited a lower superconducting transition temperature by ~ 1 K than YbC$_6$ samples prepared by vapor-phase methods ($T_c$ = 6.5 K),[14,15] possibly due to the stacking sequence disorder.

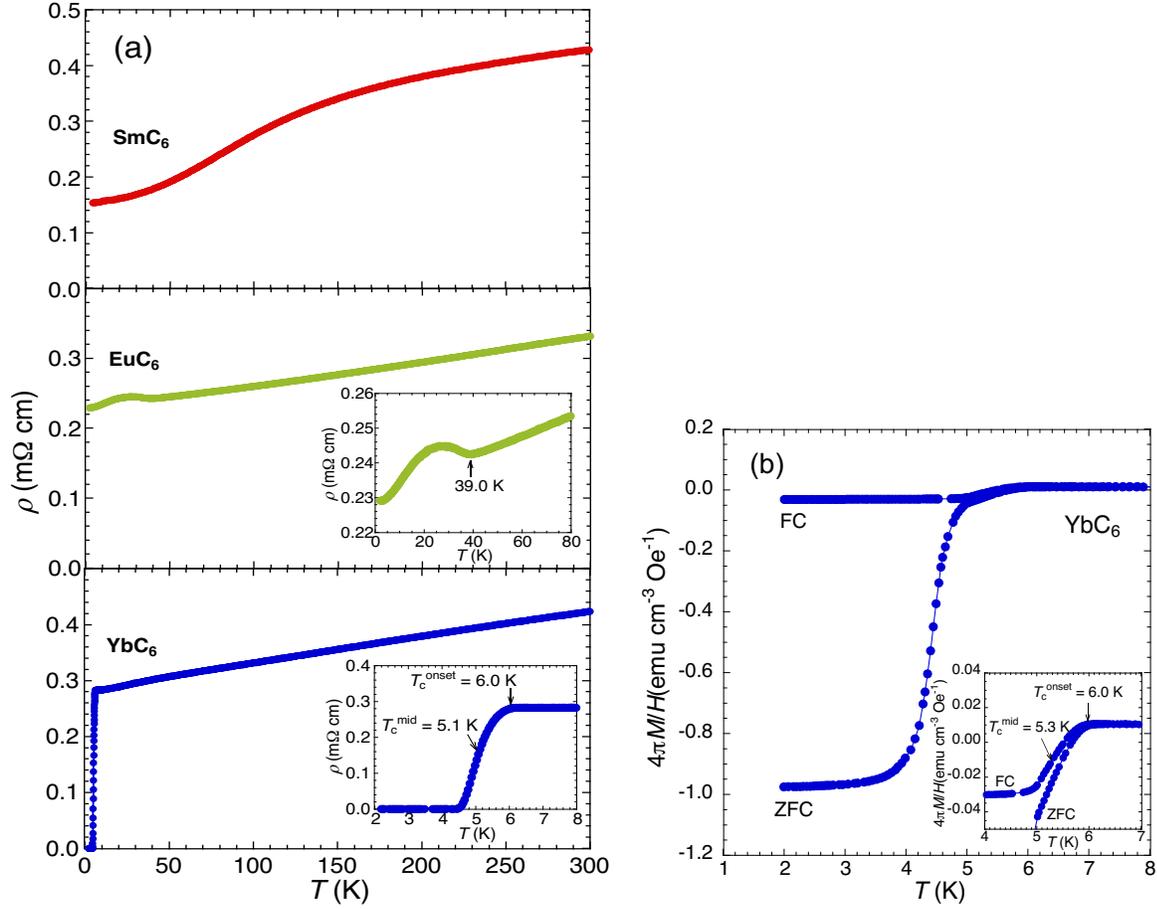

**Figure 6**. (a) $T$ dependence of $\rho$ for polycrystalline $Ln$C$_6$ ($Ln$ = Sm, Eu, and Yb) in the range of 2–300 K. The inset provides an enlarged view near the magnetic and superconducting transitions. $T_c^{mid}$ is defined as the $T$ where $\rho$ decreases to half of the value in the normal state just above $T_c$. (b) $T$ dependence of $M$ for YbC$_6$ in the range of 2–8 K. $M$ was measured in the zero-field-cooling (ZFC) and field-cooling (FC) modes in a magnetic field ($H$) of 10 Oe. $M$ was corrected by the demagnetization coefficient based on the sample geometry. The inset shows an enlarged view near $T_c$.

Figure 7a shows the shift of the resistive superconducting transition by applying $H$ of up to 14 kOe. The nearly parallel transition shift indicates the good sintering and homogeneity of the pellet. Figure 7b presents the dependence of the upper critical field ($H_{c2}$) on $T$, defined at $T_c^{onset}$, $T_c^{mid}$, and $T_c^{end}$ as marked by arrows in the figure, along with previously reported data for YbC$_6$.[14,15]

For all YbC$_6$ shown in Figure 7b, $H_{c2}$ increased linearly with lowering $T$, a common feature also observed in CaC$_6$.[7,25] However, the $H_{c2}$ in this study was enhanced by a factor of 3–5 over previously reported values for YbC$_6$ slowly synthesized by the vapor-phase method. Extrapolating $H_{c2}$ defined at $T_c^{mid}$ to 0 K yields $H_{c2}(0)$ = 16.0 kOe, as indicated in Figure 7a. The coherence length $\xi_{GL}(0)$ is derived to be 143 Å from the Ginzburg-Landau formula $H_{c2} = \Phi_0/(2\pi\xi_{GL}^2)$, where $\Phi_0$ is the flux quantum. The $\xi_{GL}(0)$ is much shorter than previously reported values of 170–450 Å.[14,15,26] The stacking sequence disorder suggested by the powder XRD analysis is possibly responsible for the short $\xi_{GL}(0)$, resulting in the enhanced $H_{c2}$. A similar trend was observed for $CaC_6$ synthesized using the Na-catalyzed method[7]; however, it was more pronounced for $YbC_6$.

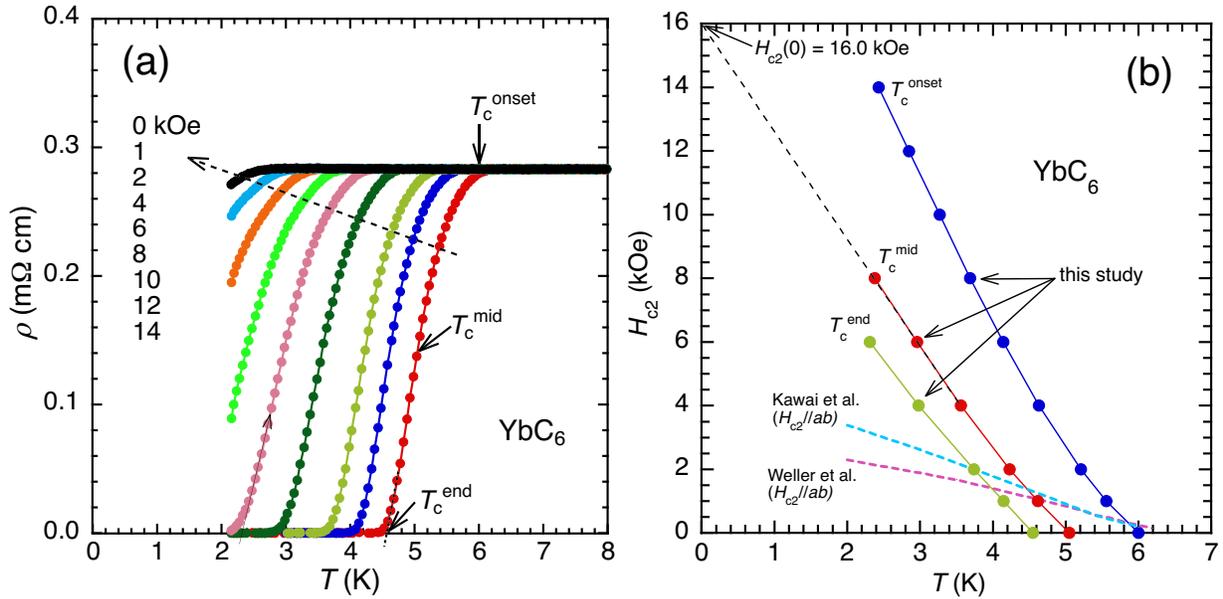

**Figure 7**. (a) Superconducting resistive transition shift in $YbC_6$ in a magnetic field from 0 to 14 kOe. The pellet is *c*-axis-oriented perpendicular to its plane, and *H* is applied to a rectangular sample cut from the pellet such that it is parallel to the ab plane (*H//ab*). (b) Dependence of $H_{c2}$ on *T*, defined at $T_c^{onset}$, $T_c^{mid}$, and $T_c^{end}$, along with previously reported data. The dashed line represents the linear extrapolation of $H_{c2}$ defined at $T_c^{mid}$.

## 4. CONCLUSIONS

The Na-catalyzed method was demonstrated to be highly effective for the synthesis of $LnC_6$ (*Ln* = Sm, Eu, and Yb). In particular, $SmC_6$ and $YbC_6$, which proved to be difficult to synthesize without the addition of Na, were rapidly formed in the presence of Na. Systematic synthesis experiments clearly captured the process by which the reaction intermediate $NaC_x$ reacted with *Ln* to form $SmC_6$ and $YbC_6$. Well-sintered pellets with reduced catalytic Na concentrations (*Ln*:Na ≈ 98:2) were shown to be feasible. $YbC_6$, located at the structural phase boundary, exhibited stacking sequence disorder because of its rapid formation catalyzed by Na, which is a possible cause of the large $H_{c2}$ enhancement. The Na-catalyzed method has versatility and is expected to be applied to more diverse intercalates in the future. Thus, this study lays the foundation for the rapid mass production of $Ln$C$_6$. The development of a more practical synthesis method for $Ln$C$_6$, along with an assessment of its potential applications, is a direction for future work.

## AUTHOR INFORMATION

**Corresponding Author**

Akira Iyo, E-mail: iyo-akira@aist.go.jp

The manuscript was written through the contributions of all authors. All authors have approved the final version of the manuscript.

**Funding Sources**

The Japan Society for the Promotion of Science (JSPS) Grants-in-Aid for Scientific Research (KAKENHI) (22K04193).

**Notes**

The authors declare no competing financial interest.

**Data Availability Statement**

The data supporting the findings of this study are available from the corresponding author upon reasonable request.

## ACKNOWLEDGMENT

This study was partly supported by the Japan Society for the Promotion of Science (JSPS) Grants-in-Aid for Scientific Research (KAKENHI) (22K04193).